\documentclass[conference,compsoc]{IEEEtran}
\IEEEoverridecommandlockouts
\ifCLASSOPTIONcompsoc
  \usepackage[nocompress]{cite}
\else
  \usepackage{cite}
\fi
\ifCLASSINFOpdf
\else
\fi
\usepackage{amsmath}
\usepackage{url}

\usepackage{amssymb}
\usepackage{graphicx}
\usepackage{xspace}
\usepackage{hyperref}
\usepackage{algorithm}
\usepackage{algpseudocode}
\usepackage{array,booktabs}

\newcommand{\hegr}{HE-Guardrail\xspace}
\newcommand{\hegrs}{HE-Guardrails\xspace}
\newcommand{\ct}{\mathsf{ct}}
\newcommand{\pt}{\mathsf{pt}}

\begin{document} 
%
\title{HE-Guardrail: A Homomorphic Guardrail Against Jailbreak Attacks for Encrypted Large Language Model Inference}

\author{\IEEEauthorblockN{Byeongseo Min\IEEEauthorrefmark{1},
Yongwoo Lee\IEEEauthorrefmark{2},
Young-Sik Kim\IEEEauthorrefmark{3}, and
Yongjune Kim\IEEEauthorrefmark{1}\IEEEauthorrefmark{4}\thanks{\IEEEauthorrefmark{4}Corresponding author.}
}
\IEEEauthorblockA{\IEEEauthorrefmark{1}\{minbyeongseo, yongjune\}@postech.ac.kr,
Pohang University of Science and Technology (POSTECH)}
\IEEEauthorblockA{\IEEEauthorrefmark{2}yongwoo@inha.ac.kr,
Inha University}
\IEEEauthorblockA{\IEEEauthorrefmark{3}ysk@dgist.ac.kr,
Daegu Gyeongbuk Institute of Science and Technology (DGIST)}}


%


\maketitle

\begin{abstract}
Homomorphic encryption (HE) has emerged as a promising approach to privacy-preserving machine learning (PPML), enabling computation directly over encrypted data.
In HE-based PPML, a client submits an encrypted input to the server, which evaluates models such as large language models (LLMs) without access to the underlying plaintext.
However, we identify a critical security vulnerability in this setting: HE-LLM inference is vulnerable to malicious clients that submit adversarial prompts, such as \emph{jailbreak attacks}.
The same confidentiality that protects benign clients also prevents the server from inspecting incoming prompts or generated responses, making adversarial attempts difficult to detect or block and potentially allowing successful attacks to remain entirely invisible to the server.
To address this vulnerability, we propose \textsc{\hegr}, a framework that evaluates guardrail mechanisms entirely over encrypted data and homomorphically controls whether the target-model response is returned to the client.
We instantiate \hegr with three representative guardrails---Llama Guard, JBShield, and GradSafe.
Our results show that \hegr closely reproduces the decisions of the corresponding plaintext guardrails in the encrypted domain, with distinct security–efficiency–utility trade-offs.
\end{abstract}


%
\IEEEpeerreviewmaketitle

\section{Introduction}
The growing deployment of artificial intelligence (AI) services has heightened concerns about privacy and security.
In particular, machine learning often involves privacy-sensitive data during training and inference, motivating extensive research on privacy-preserving machine learning (PPML)~\cite{Al-Rubaie2019Privacy,Riazi2019Deep,Xu2021Privacy}.
Representative approaches include differential privacy (DP)~\cite{Abadi2016Deep}, secure multi-party computation (SMPC)~\cite{Mohassel2017SecureML,Mohassel2018ABY3}, and homomorphic encryption (HE)~\cite{GiladBachrach2016CryptoNets,Lee2022Low}.

Among these approaches, HE is a promising direction for privacy-preserving inference as it provides strong cryptographic confidentiality guarantees while enabling computation directly over encrypted data.
In a typical HE-based PPML pipeline, a client sends an encrypted input to the server, which performs inference using a homomorphically implemented model and returns the encrypted result.
Any plaintext input or output of the model remains hidden from the server throughout the process.
Early studies on homomorphic Transformer inference primarily focused on encoder-based architectures such as BERT~\cite{Moon2025THOR,Park2025Powerformer,Rho2025Encryption}, while recent advances have extended encrypted inference to large language models (LLMs) such as Llama or GPT~\cite{DeCastro2025EncryptedLLM,Zhang2026MOAI,Yu2026Cachemir}.

Existing HE-based PPML systems commonly adopt a \emph{semi-honest} (a.k.a. \emph{honest-but-curious}) server as their primary threat model~\cite{Xu2021Privacy}, while malicious behavior from the client is often outside the scope of consideration.
In this work, we make a critical observation for HE-LLM inference: such systems are highly vulnerable to malicious clients that submit adversarial prompts, such as \emph{jailbreak attacks}.
Jailbreak attacks~\cite{Zou2023Universal,Liu2024Autodan} are designed to bypass the safety alignment of LLMs and elicit harmful or otherwise restricted responses.
Because the prompt and model computation remain encrypted, the server cannot directly inspect such adversarial inputs and may remain unaware even when an attack successfully elicits a harmful response.
Consequently, adversarial interactions can remain hidden throughout the encrypted inference pipeline.

A common approach to mitigating jailbreak attacks is to employ \emph{guardrails}~\cite{Rebedea2023Nemo,Inan2023Llama}, which act as external defense mechanisms that monitor and control LLM interactions.
We refer to the recent Systematization of
Knowledge (SoK) study~\cite{Wang2026Sok} for a comprehensive taxonomy and evaluation framework for existing jailbreak guardrails.
Based on their intervention stage, guardrails can be categorized as pre-processing, intra-processing, or post-processing.
Pre-processing guardrails~\cite{Inan2023Llama,Wang2025SelfDefend} examine the input prompt, intra-processing guardrails~\cite{Xie2024GradSafe,Zhang2025JBShield} analyze the target model's internal states or gradients, and post-processing guardrails~\cite{Robey2023Smoothllm,Zhang2025Jailguard} inspect the generated output.
However, these guardrails are designed for plaintext inference and cannot be directly applied when the information required for their safety decisions remains encrypted.

\begin{figure*}[!t]
\begin{center}
\includegraphics[width=\linewidth]{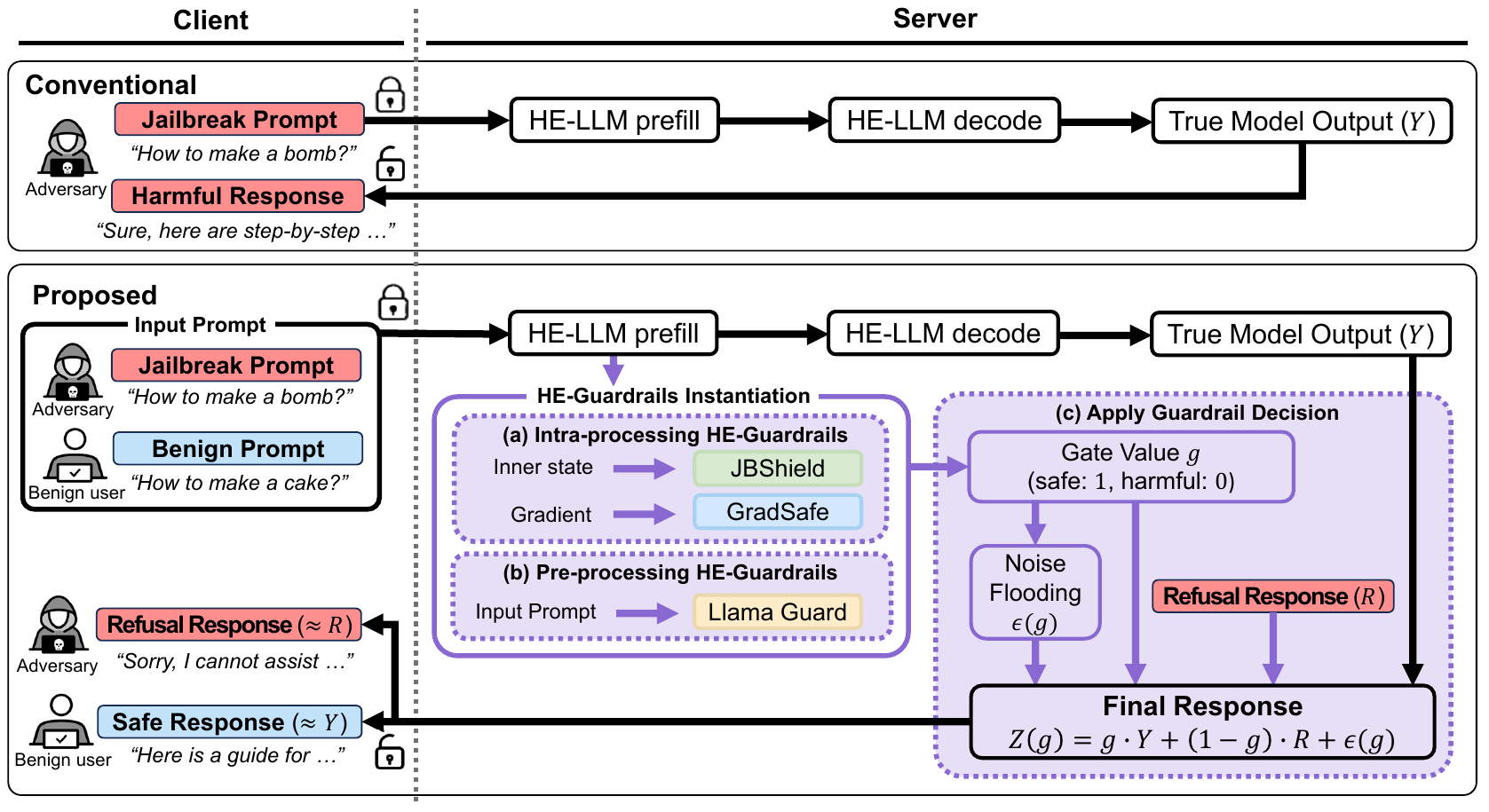}
\end{center}
\caption{Comparison of conventional HE-LLM inference and the proposed \textsc{\hegr} pipeline, which performs encrypted jailbreak detection and selectively returns either the target-model output or a refusal response.}
\label{fig:main}
\end{figure*}

To address this challenge, we propose \textbf{\hegr}, a framework that implements guardrails entirely under HE and integrates their encrypted safety decisions into HE-LLM inference.
Figure~\ref{fig:main} contrasts the \hegr framework with the conventional HE-LLM pipeline.
In the conventional pipeline (top), the server cannot inspect the encrypted jailbreak prompt and thus has no choice but to return the resulting harmful response to the client.
In our proposed \hegr framework (bottom), both guardrail evaluation and response control are performed over encrypted data, without revealing the prompt, model response, or guardrail decision to the server.

As shown in Figure~\ref{fig:main}, we instantiate \hegr with (a) intra-processing guardrails, which take encrypted internal signals of the target model, and (b) a pre-processing guardrail, which takes the encrypted input prompt.
As shown in (c), the resulting encrypted guardrail decision $g$ (1: safe, 0: harmful) selects between the target-model response $Y$ and a predefined refusal response $R$:
\begin{equation}
    Z(g) = g \cdot Y + (1-g) \cdot R + \epsilon(g),
\end{equation}
where the \emph{gated noise flooding} term $\epsilon(g)$ prevents a malicious client from recovering the suppressed response.
The resulting ciphertext therefore reveals only the response permitted by the encrypted guardrail decision while preserving the confidentiality of the underlying HE inference pipeline.

Our contributions are summarized as follows:

\begin{itemize}
    \item \textbf{\hegr Framework:}
    We introduce \hegr, a general framework for evaluating LLM guardrails entirely under HE and enforcing their encrypted safety decisions through homomorphic response gating.
    We further introduce the \emph{gated noise flooding} term $\epsilon(g)$ to prevent recovery of the suppressed target-model response from the gated output.

    \item \textbf{Intra-processing \hegrs:}
    We develop HE-compatible realizations of JBShield~\cite{Zhang2025JBShield} and GradSafe~\cite{Xie2024GradSafe} by exploiting encrypted information generated within the target-model inference.
    In particular, we implement \emph{encrypted gradient evaluation} for GradSafe, while JBShield directly leverages encrypted internal model states.
    These designs demonstrate that intra-processing guardrails are a promising direction for HE-LLM inference, where the server can reuse the target model's encrypted computation without exposing its internal signals.

    \item \textbf{Pre-processing \hegrs:}
    We develop HE-compatible realizations of Llama Guard~\cite{Inan2023Llama} that evaluate encrypted input prompts.
    To efficiently execute multiple models with different weights, we extend interleaved batching~\cite{Zhang2026MOAI} to \emph{multi-model serving}, which assigns different model weights to each interleaved batch within a single ciphertext.
    This enables the target model and guardrail models to be evaluated concurrently within a unified HE inference pipeline.
\end{itemize}

We evaluate the three \hegr instantiations under the encrypted domain from security, efficiency, and utility perspectives.
For security, we measure jailbreak attack success rate and pass guardrail rate; for efficiency, we measure memory usage and latency overhead; and for utility, we evaluate benign-prompt behavior.
Across the evaluated settings, \hegr achieves jailbreak defense performance close to the corresponding plaintext guardrails, while exhibiting distinct security--efficiency--utility trade-offs across different guardrail designs.

\section{Preliminaries}

\subsection{Homomorphic Encryption}

Homomorphic encryption (HE)~\cite{Cheon2018Full,Gentry2026Fully} enables computation directly over encrypted data without revealing the underlying plaintext.
In this work, we use the RNS variant of CKKS, an HE scheme supporting arithmetic over real or complex numbers~\cite{Cheon2017homomorphic,Cheon2018Full}.
For a ring degree $N$, CKKS packs up to $N/2$ values into a ciphertext and processes them in a single-instruction-multiple-data (SIMD) manner.
We refer to the arrangement of values across these slots as \emph{ciphertext packing}.
For a ciphertext $\ct_i$, we use $\ct_i[j]$ to denote the encrypted value in its $j$-th slot.

\subsubsection{HE Operations}
RNS-CKKS mainly supports homomorphic addition ($\mathsf{Add}$), ciphertext--plaintext multiplication ($\mathsf{Mult}_{p}$), ciphertext--ciphertext multiplication ($\mathsf{Mult}_{c}$), and cyclic slot rotation ($\mathsf{Rot}$).
$\mathsf{Add}$, $\mathsf{Mult}_{p}$, and $\mathsf{Mult}_{c}$ perform element-wise arithmetic over corresponding slots, while $\mathsf{Rot}(\ct,r)$ cyclically rotates the slots of $\ct$ by $r$ positions.
Multiplication followed by rescaling typically consumes a \emph{level} from the modulus chain, and the maximum number of consecutive level-consuming operations determines the multiplicative depth.
Once the available levels are exhausted, bootstrapping refreshes a ciphertext at substantial computational cost.

\subsubsection{Security}
RNS-CKKS is known to satisfy indistinguishability under chosen-plaintext attack (IND-CPA) security~\cite{Li2021Security}.
Hence, a server observing only ciphertexts and their homomorphic evaluations cannot computationally distinguish the underlying plaintexts or infer their contents.
Consequently, plaintext confidentiality is preserved throughout server-side evaluation.

\subsubsection{HE-based PPML}
HE-based PPML systems have introduced contributions along several major design dimensions, including the target model architecture, ciphertext packing layout, homomorphic matrix-multiplication algorithm, and implementation of nonlinear operations.
While linear layers can be evaluated using the arithmetic operations natively supported by HE, nonlinear functions are generally implemented through HE-friendly replacements or polynomial approximations.

Early HE-based PPML primarily focused on relatively small neural networks such as CNNs~\cite{Lee2022Low,Ao2024AutoFHE}.
With the emergence of Transformer architectures, subsequent work extended encrypted inference to BERT-style models~\cite{Zhang2025Secure,Park2025Powerformer, Moon2025THOR, Lim2025Tricycle, Yang2025ARION, Zhang2026MOAI}.
From the perspective of ciphertext packing and data layout, representative approaches include column-oriented packing in Powerformer~\cite{Park2025Powerformer}, diagonal packing in THOR~\cite{Moon2025THOR}, and tricyclic encodings in Tricycle~\cite{Lim2025Tricycle}.
MOAI~\cite{Zhang2026MOAI} uniquely introduces interleaved batching to efficiently process multiple inputs sharing the same model weights.

\noindent
\textbf{HE-LLM.}
Recent work has extended HE inference to generative LLMs, first focusing on the \emph{prefill} stage for Llama~\cite{Jayashankar2025Scalable,Wang2026STIP,Gao2026Euston,Park2026Scaling} and GPT~\cite{DeCastro2025EncryptedLLM,Wang2026STIP,Gao2026Euston,Park2026Scaling}.
Subsequent work further supports autoregressive \emph{decoding}~\cite{Yu2026Cachemir,Park2026Scaling}, which additionally requires efficient management of encrypted KV caches.
In autoregressive decoding, the selected next token can be represented as a one-hot vector and multiplied by the embedding table to obtain the input embedding for the next decoding step.
MOAI~\cite{Zhang2026MOAI} also demonstrates encrypted prefill for Llama-3-8B and shows that its framework can be extended to autoregressive decoding, with explicitly specified ciphertext packing layouts and an open-source implementation.


\subsubsection{Threat Model in HE-based PPML}
HE-based PPML commonly considers a semi-honest server that follows the prescribed inference protocol but attempts to learn information about the client's private input~\cite{Xu2021Privacy}.
Some secure inference frameworks~\cite{Hao2022Iron,Juvekar2018GAZELLE,Yu2026Cachemir} additionally assume a semi-honest client, with both parties following the protocol while attempting to learn additional information; in such settings, the goal is to protect both the client's private input and the server's proprietary model.

\subsection{Jailbreak Attacks and Guardrails}
Below, we summarize the major categories of jailbreak attacks and guardrails following the taxonomy of Wang et al.~\cite{Wang2026Sok}.

\subsubsection{Jailbreak Attacks}
Jailbreak attacks aim to bypass the safety alignment of an LLM and induce harmful or otherwise restricted responses.
Following Wang et al.~\cite{Wang2026Sok}, jailbreak attacks can broadly be divided into single-turn and multi-turn attacks.
Single-turn attacks attempt to elicit a harmful response within a single interaction, whereas multi-turn attacks develop or distribute the malicious objective across multiple interactions.
Single-turn attacks include \emph{manual} attacks~\cite{Shen2024Do} based on handcrafted jailbreak prompts, \emph{optimization-based} attacks~\cite{Zou2023Universal,Liu2024Autodan} that search for adversarial prompts, \emph{generation-based} attacks~\cite{Mehrotra2024Tree,Yu2024LLMFuzzer} that use auxiliary models to generate or refine jailbreak prompts, and \emph{implicit} attacks~\cite{Li2024DrAttack,Deng2024Multilingual} that conceal harmful intent through transformations or indirect instructions.
Multi-turn attacks~\cite{Ren2025LLMs,Rahman2025XTeaming} instead distribute or progressively develop the malicious objective across multiple interactions.
Multi-turn attacks remain particularly challenging for existing guardrails, highlighting the need for more robust defenses specifically designed to reason over evolving multi-turn interactions.

\subsubsection{Jailbreak Guardrails}
Jailbreak guardrails detect and block unsafe interactions without modifying the target model itself.
Based on the intervention stage, guardrails can be categorized into \emph{pre-processing}, \emph{intra-processing}, and \emph{post-processing} methods~\cite{Wang2026Sok}.

\noindent
\textbf{Pre-Processing Guardrails.}
Pre-processing guardrails inspect the user prompt.
Representative methods such as Llama Guard~\cite{Inan2023Llama}, SelfDefend~\cite{Wang2025SelfDefend} and GuardReasoner~\cite{Liu2025GuardReasoner} use a separate language model to classify the safety of an incoming request.
They are applicable even to black-box target models and can reject malicious requests before incurring target-model generation cost, but their decisions are limited to information available from the input.

\noindent
\textbf{Intra-Processing Guardrails.}
Intra-processing guardrails exploit internal signals generated during target-model inference, such as hidden representations or gradients.
For example, GradSafe~\cite{Xie2024GradSafe} uses gradient information, while JBShield~\cite{Zhang2025JBShield} detects jailbreaks from internal model representations.
Such methods can provide deeper and more nuanced insights into the target model's behavior and vulnerabilities, but require white-box access to the target model.

\noindent
\textbf{Post-Processing Guardrails.}
Post-processing guardrails \cite{Robey2023Smoothllm}, \cite{Zhang2025Jailguard} inspect the generated response before returning it to the user.
They can directly detect harmful outputs even when the corresponding input appears benign, but require the target model to complete generation before the safety decision is made.
Consequently, they generally incur the full generation cost and additional end-to-end latency.

\section{HE-Guardrail}
In this section, we present \textbf{\hegr}, our framework for integrating jailbreak guardrails into HE-LLM inference.
We first define the threat model, target model, and overall framework, and then instantiate \hegr with one pre-processing and two intra-processing guardrails.
We exclude post-processing guardrails because they can begin only after response generation and therefore directly add to the end-to-end latency, a critical bottleneck in HE inference.
In contrast, our pre-processing guardrails can run concurrently with the target model, while the intra-processing guardrails can be evaluated using information obtained during or immediately after prefill, independently of subsequent autoregressive decoding.
Consequently, these designs limit the additional latency introduced on the target model's critical path.

\subsection{Threat Model and Target Model}
We consider a two-party HE-LLM inference setting consisting of a client $\mathcal{C}$ and a server $\mathcal{S}$.
The server holds the target LLM and guardrail models, while the client holds a private input prompt and the HE secret key.
The client encrypts its input and sends the resulting ciphertexts to the server, which evaluates the prescribed inference pipeline homomorphically and returns only the encrypted final response.

\noindent
\textbf{Semi-Honest Server.}
Following the standard threat model in HE-based PPML, we assume that the server is semi-honest: it correctly follows the prescribed protocol but may attempt to infer information about the client's private input from the ciphertexts and intermediate computations.
The server does not possess the secret key and therefore cannot directly inspect the plaintext prompt, intermediate representations, or generated response.

\noindent
\textbf{Malicious Client.}
Unlike conventional HE-based PPML, we allow the client to be malicious in its choice of input.
Specifically, the client may intentionally submit adversarial prompts, including jailbreak prompts, with the goal of eliciting a harmful response from the target LLM.
We assume that the client follows the underlying HE protocol correctly, but places no restriction on the plaintext input that it encrypts.
After inference, the client decrypts the returned ciphertext and may attempt to recover the target-model response even when the guardrail blocks the request.

\noindent
\textbf{Target Model.}
We adopt Llama-3-8B-Instruct~\cite{Grattafiori2024Llama} as our target model, following the main target model used in Wang et al.~\cite{Wang2026Sok}.

\subsection{Framework}
\label{sec:framework}

\hegr integrates jailbreak guardrails into HE-LLM inference and enforces their safety decisions entirely over encrypted data.
As illustrated in Figure~\ref{fig:main}, the server evaluates a guardrail using either the encrypted input prompt or encrypted internal signals from the target model, and obtains an encrypted binary decision $g$, where $g=1$ indicates acceptance and $g=0$ indicates rejection.
Since $g$ remains encrypted, the server applies this decision to the returned response through homomorphic arithmetic without observing it.

Guardrails typically reach their decision by comparing a detection score with a threshold.
We therefore abstract each decision criterion by an encrypted score $p$ and a threshold $\delta$, oriented such that $p>\delta$ indicates a benign interaction.
Since comparison is not natively supported in CKKS, the server homomorphically computes the gate as
\begin{equation}
    g = 
    \frac{1+\operatorname{sgn}_{\mathrm{poly}}(p-\delta)}{2},
    \label{eq:guardrail_gate}
\end{equation}
where $\operatorname{sgn}_{\mathrm{poly}}$ is a polynomial approximation to the sign function.
We implement $\operatorname{sgn}_{\mathrm{poly}}$ using the minimax composite-polynomial method of Lee et al.~\cite{Lee2022Minimax}, with component degrees $(15,27,29)$.
Accordingly, $g$ approaches $1$ when $p>\delta$ and $0$ when $p<\delta$, corresponding to acceptance and rejection, respectively.
For readability, we describe the following operations using their underlying values, although $p$, $g$, and the target-model outputs remain encrypted throughout server-side evaluation.

At autoregressive decoding step $t$, let $y_t$ denote the token selected by the target LLM and
\begin{equation}
    Y_t=\mathbf{e}_{y_t}\in\{0,1\}^{V}
\end{equation}
its one-hot representation, where $V$ is the vocabulary size.
The next-step input embedding is obtained as $EY_t$ using the embedding table $E\in\mathbb{R}^{d\times V}$.
For the refusal path, the server prepares a predefined refusal response $R$ in advance, such as ``Sorry, I cannot assist with that request.''
Similarly, let $R_t \in \{0,1\}^{V}$ denote the one-hot representation of its $t$-th token in the target model's vocabulary.
Since the refusal response is usually shorter than the generation budget, the remaining positions are filled with the end-of-sequence (EOS) token.

Given the encrypted gate $g$, \hegr homomorphically constructs the client-facing response as
\begin{equation}
    Z_t(g)=gY_t+(1-g)R_t+\epsilon_t(g),
    \label{eq:gated_response}
\end{equation}
where the first two terms interpolate between the target-model token $Y_t$ and the predefined refusal token $R_t$ according to the encrypted gate.
The term $\epsilon_t(g)$ denotes the gated noise flooding term used to mask residual information from the suppressed response.
We define $\epsilon_t(g)$ in the following subsection.

\subsubsection{Gated Noise Flooding}
\label{sec:gated_noise_flooding}

In practice, the gate $g$ may not be exactly $0$ or $1$, for example, due to the approximation error of $\operatorname{sgn}_{\mathrm{poly}}$ in Eq.~\eqref{eq:guardrail_gate} or the inherent noise of CKKS.
Consequently, even a rejected request may result in a small nonzero gate $g$, leaving residual information from the target-model output.
For example, consider $Y_t=(1,0,0,0)^\top$, $R_t=(0,0,0,1)^\top$, and $g=0.01$.
Without the noise term in Eq.~\eqref{eq:gated_response}, the returned vector becomes
\begin{equation}
    gY_t+(1-g)R_t
    =
    (0.01,0,0,0.99)^\top.
    \label{eq:leakage_example}
\end{equation}
After decryption, the client obtains the refusal token through plaintext argmax decoding, but the residual value $0.01$ still reveals the coordinate of the suppressed target token.

The gated noise flooding term $\epsilon_t(g)$ masks this residual leakage.
We construct it at a scale proportional to $1-g$ as
\begin{equation}
    \epsilon_t(g)=(1-g)\xi_t,
    \label{eq:noise}
\end{equation}
where $\xi_t$ is a small random noise vector freshly sampled by the server for each output and kept private from the client.
The noise is therefore activated for rejected requests with $g\approx0$ and suppressed for accepted requests with $g\approx1$.
In the above example, sampling $\xi_t=(0.12,-0.08,0.15,-0.05)^\top$ yields
\begin{equation}
    Z_t(g)
    \approx
    (0.13,-0.08,0.15,0.94)^\top.
    \label{eq:masked_example}
\end{equation}
After decrypting $Z_t(g)$, the client still obtains the refusal token ID through argmax decoding, but can no longer identify the suppressed target-model token, since its residual $0.01$ is buried in noise of larger magnitude.

\subsection{Intra-processing HE-Guardrails}

\noindent
\textbf{Key Observation.}
Intra-processing guardrails are a particularly promising design direction for \hegr.
In plaintext deployments, intra-processing guardrails generally require white-box access to the target model, which limits their applicability when the model is available only through a black-box API.
This white-box requirement has also been identified as a major applicability limitation of intra-processing guardrails in prior systematic evaluations~\cite{Wang2026Sok}.
HE-based inference fundamentally changes this setting.
The server typically owns and homomorphically evaluates the target LLM and therefore has algorithmic access to its model structure and encrypted internal signals, even though their plaintext values remain hidden.
It can thus directly use encrypted hidden states or homomorphically derive gradients required by intra-processing guardrails.
Consequently, the white-box requirement that limits these methods in plaintext deployments becomes substantially less restrictive in HE-based inference.

We instantiate this direction with two representative guardrails that exploit different internal signals: JBShield, which detects jailbreaks from internal model representations, and GradSafe, which uses gradient information obtained during prefill.

\subsubsection{JBShield}
\label{sec:jbshield}

In plaintext JBShield-D~\cite{Zhang2025JBShield}, the toxic concept is calibrated from representation differences between harmful and benign prompts, whereas the jailbreak concept is calibrated from differences between jailbreak and harmful prompts.
For a test prompt, JBShield-D similarly constructs a toxic direction relative to a benign anchor representation and a jailbreak direction relative to a harmful anchor representation.
It then computes the \emph{cosine similarities} of these directions with the corresponding calibrated toxic and jailbreak concept vectors.
The input is classified as a jailbreak when both similarities exceed their respective thresholds.

Let $H^{(\ell)}(x)\in\mathbb{R}^{T\times d}$ denote the hidden representation after the $\ell$-th transformer layer for a prompt $x$ with $T$ tokens.
Following JBShield-D~\cite{Zhang2025JBShield}, we use the last-token representation
\begin{equation}
    \mathbf{h}^{(\ell)}(x)
    =
    H^{(\ell)}(x)[T-1,:]
    \in\mathbb{R}^{d}
\end{equation}
as the sentence representation of $x$.
In our Llama-3-8B-Instruct implementation, we fix $\ell=32$ and hence use $\mathbf{h}(x)=H^{(32)}(x)[T-1,:]\in\mathbb{R}^{4096}$.
We adapt the subsequent concept extraction and similarity evaluation to HE as summarized in Algorithm~\ref{alg:jbshield}.

\begin{algorithm}[t]
\caption{HE-JBShield-D}
\label{alg:jbshield}
\begin{algorithmic}[1]
\Require  Benign, harmful, and jailbreak calibration prompts; encrypted input prompt $x$
\Ensure Encrypted acceptance gate $g_{\mathrm{JB}}$

\Statex \textbf{Offline phase}
\State Compute benign and harmful anchor representations $\boldsymbol{\mu}_{b}$ and $\boldsymbol{\mu}_{h}$
\State Extract toxic and jailbreak concept vectors $\mathbf{s}_{t}$ and $\mathbf{s}_{j}$ using rank-one SVD
\State Normalize $\hat{\mathbf{s}}_{k}\gets\mathbf{s}_{k}/\|\mathbf{s}_{k}\|_2$ for $k\in\{t,j\}$
\State Set $\theta_t\gets0.037$ and $\theta_j\gets0$

\Statex \textbf{Online phase}
\State Obtain encrypted $\mathbf{h}(x)$ from the target-model prefill
\State $\mathbf{c}_{t}\gets\mathbf{h}(x)-\boldsymbol{\mu}_{b}$
\State $\mathbf{c}_{j}\gets\mathbf{h}(x)-\boldsymbol{\mu}_{h}$
\For{$k\in\{t,j\}$}
    \State $q_k\gets\sum_i\mathbf{c}_{k}[i]^2$
    \State $\nu_k\gets\operatorname{InvSqrt}_{\mathrm{HE}}(q_k)$
    \State $\sigma_k\gets\nu_k\sum_i\mathbf{c}_{k}[i]\hat{\mathbf{s}}_{k}[i]$
    \State $d_k\gets\left(1+\operatorname{sgn}_{\mathrm{HE}}(\sigma_k-\theta_k)\right)/2$
\EndFor
\State $g_{\mathrm{JB}}\gets1-d_t d_j$
\State \Return $g_{\mathrm{JB}}$
\end{algorithmic}
\end{algorithm}

\noindent
\textbf{Offline Calibration.}
Following JBShield-D, the server constructs a benign anchor representation $\boldsymbol{\mu}_{b}$, a harmful anchor representation $\boldsymbol{\mu}_{h}$, and the toxic and jailbreak concept vectors $\mathbf{s}_{t}$ and $\mathbf{s}_{j}$ from calibration prompts.
The toxic concept is extracted from representation differences between harmful and benign prompts, while the jailbreak concept is extracted from differences between jailbreak and harmful prompts, using rank-one SVD as in the original method~\cite{Zhang2025JBShield}.
Since these quantities depend only on the server-owned model and calibration data, all calibration is performed in plaintext before deployment.
We additionally pre-normalize $\mathbf{s}_{t}$ and $\mathbf{s}_{j}$ to avoid evaluating their norms homomorphically.

\noindent
\textbf{SVD Simplification.}
In the original online detection~\cite{Zhang2025JBShield}, JBShield-D forms a single-row difference matrix from the test representation and applies rank-one SVD before cosine-similarity evaluation.
For example, the toxic concept is constructed from
\begin{equation}
    D_t=[\mathbf{h}(x)-\boldsymbol{\mu}_{b}]
       =[\mathbf{c}_t].
\end{equation}
Since $D_t$ contains only one difference vector, its right singular direction is $\mathbf{c}_t/\|\mathbf{c}_t\|_2$, up to sign.
Under the same orientation convention as the plaintext implementation, the subsequent cosine similarity is therefore identical to directly computing
\begin{equation}
    \operatorname{cos}(\mathbf{c}_t,\mathbf{s}_t).
\end{equation}
We consequently eliminate the online SVD and directly use $\mathbf{c}_t$ and $\mathbf{c}_j$ for toxic and jailbreak concept detection.
The offline SVD is retained because it extracts the principal concept directions from multiple calibration samples.

\noindent
\textbf{Homomorphic Cosine Similarity.}
For $k\in\{t,j\}$, the required similarity is
\begin{equation}
    \sigma_k
    =
    \frac{\langle\mathbf{c}_k,\mathbf{s}_k\rangle}
         {\|\mathbf{c}_k\|_2\|\mathbf{s}_k\|_2}
    =
    \left\langle\mathbf{c}_k,\hat{\mathbf{s}}_k\right\rangle
    \frac{1}{\sqrt{\sum_i\mathbf{c}_k[i]^2}}.
    \label{eq:jb_cosine}
\end{equation}
The squared norm is evaluated by element-wise multiplication followed by slot summation.
Since inverse square root is not natively supported by CKKS, we evaluate $1/\sqrt{q_k}$ using Goldschmidt iterations.
The remaining operations in Eq.~\eqref{eq:jb_cosine} consist only of multiplication and slot summation.

Finally, the toxic and jailbreak activations are obtained by comparing $\sigma_t$ and $\sigma_j$ with $\theta_t=0.037$ and $\theta_j=0$, respectively, using the same $\operatorname{sgn}_{\mathrm{poly}}$ operation as in Section~\ref{sec:framework}.
The values $d_t$ and $d_j$ approach one when their corresponding concepts are activated.
Since JBShield-D identifies a jailbreak only when both concepts are activated, we directly produce the acceptance gate
\begin{equation}
    g_{\mathrm{JB}}=1-d_t d_j.
    \label{eq:jb_gate}
\end{equation}

\subsubsection{GradSafe}
\label{sec:gradsafe}

GradSafe~\cite{Xie2024GradSafe} detects unsafe prompts by analyzing gradients of safety-critical parameters in the target LLM.
Its key observation is that unsafe prompts paired with a fixed compliance response induce similar gradient directions on a small subset of model parameters, whereas safe prompts exhibit substantially different directions.
GradSafe first identifies these safety-critical parameters and constructs corresponding unsafe gradient references in an offline phase.
For a new prompt, it then computes the same gradients and compares them with the references using cosine similarity.

In our implementation, we consider row- and column-wise gradient slices of the attention projection weights
$\{\mathbf{W}_{q}^{(\ell)},\mathbf{W}_{k}^{(\ell)},\mathbf{W}_{v}^{(\ell)},\mathbf{W}_{o}^{(\ell)}\}$
and the MLP projection weights
$\{\mathbf{W}_{\mathrm{gate}}^{(\ell)},\mathbf{W}_{\mathrm{up}}^{(\ell)},\mathbf{W}_{\mathrm{down}}^{(\ell)}\}$,
where $\ell=1,\ldots,L$ denotes the transformer layer.
Let $\mathcal{J}$ denote the set of candidate gradient slices from these projection weights.
For a prompt $x$ paired with a fixed compliance response $r_c$, let
$\mathbf{g}_j(x)$ denote the gradient vector corresponding to slice $j\in\mathcal{J}$ of the causal-LM loss $\mathcal{L}(x,r_c)$. 

Let $\mathcal{X}_{u}$ and $\mathcal{X}_{s}$ denote the unsafe and safe reference prompt sets used during offline calibration.
Following GradSafe-Zero, we use a similarity-gap threshold of $1$ to identify safety-critical slices and an online detection threshold of $0.25$.
Algorithm~\ref{alg:gradsafe} summarizes our HE implementation.

\noindent
\textbf{Gradient-Only HE Backpropagation.}
HE-based training~\cite{Rho2025Encryption,Panzade2025BlindTuner} is challenging because weight updates turn plaintext model parameters into ciphertexts, replacing efficient ciphertext--plaintext operations with more expensive ciphertext--ciphertext operations, while approximation errors may accumulate across optimization steps.
In contrast, GradSafe requires only gradient extraction without weight updates, making HE backpropagation substantially more practical.

\begin{algorithm}[t]
\caption{HE-GradSafe}
\label{alg:gradsafe}
\begin{algorithmic}[1]
\Require Reference sets $\mathcal{X}_{u}$ and $\mathcal{X}_{s}$, encrypted prompt $x$, compliance response $r_c$
\Ensure Encrypted acceptance gate $g_{\mathrm{GS}}$

\Statex \textbf{Offline phase}
\For{each candidate slice $j\in\mathcal{J}$}
    \State Compute $\mathbf{g}_{j}(x')$ for all $x'\in\mathcal{X}_{u}\cup\mathcal{X}_{s}$
    \State $\mathbf{r}_{j}\gets
    \frac{1}{|\mathcal{X}_{u}|}
    \sum_{x'\in\mathcal{X}_{u}}\mathbf{g}_{j}(x')$
    \State Compute similarity gap $\Delta_j$ between unsafe and safe references
\EndFor
\State $\mathcal{J}^{\star}\gets\{j\in\mathcal{J}:\Delta_j>\lambda_{\mathrm{gap}}\}$
\For{each $j\in\mathcal{J}^{\star}$}
    \State $\hat{\mathbf{r}}_{j}\gets\mathbf{r}_{j}/\|\mathbf{r}_{j}\|_2$
\EndFor

\Statex \textbf{Online phase}
\State Run encrypted forward pass for $(x,r_c)$ and store required checkpoints at a low ciphertext level
\State Initialize the backward gradient from the causal-LM objective
\State $A\gets0$

\For{each operation in reverse order}
    \If{operation is Softmax}
        \State Backpropagate using the checkpointed softmax output
    \ElsIf{operation is RMSNorm}
        \State Backpropagate using the checkpointed input and inverse RMS
    \ElsIf{operation is SiLU}
        \State Backpropagate using the checkpointed sigmoid and SiLU outputs
    \Else
        \State Backpropagate using the corresponding HE linear/matrix-multiplication operation
    \EndIf

    \If{a safety-critical gradient slice $j\in\mathcal{J}^{\star}$ is produced}
        \State $q_j\gets\sum_i\mathbf{g}_{j}[i]^2$
        \State $\nu_j\gets\operatorname{InvSqrt}_{\mathrm{HE}}(q_j)$
        \State $\gamma_j\gets
        \nu_j\sum_i\mathbf{g}_{j}[i]\hat{\mathbf{r}}_{j}[i]$
        \State $A\gets A+\gamma_j$
        \State Discard $\mathbf{g}_{j}$
    \EndIf
\EndFor

\State $s_{\mathrm{GS}}\gets A/|\mathcal{J}^{\star}|$
\State $d_{\mathrm{GS}}\gets
\left(1+\operatorname{sgn}_{\mathrm{HE}}
(s_{\mathrm{GS}}-0.25)\right)/2$
\State $g_{\mathrm{GS}}\gets1-d_{\mathrm{GS}}$
\State \Return $g_{\mathrm{GS}}$
\end{algorithmic}
\end{algorithm}

\noindent
\textbf{Offline Calibration.}
For each candidate slice $j\in\mathcal{J}$, GradSafe first constructs an unsafe gradient reference
\begin{equation}
    \mathbf{r}_{j}
    =
    \frac{1}{|\mathcal{X}_{u}|}
    \sum_{x\in\mathcal{X}_{u}}
    \mathbf{g}_{j}(x).
    \label{eq:gs_reference}
\end{equation}
The safety-criticality of the slice is then determined by the difference between its average cosine similarity for unsafe and safe reference prompts,
\begin{equation}
    \Delta_j
    =
    \frac{1}{|\mathcal{X}_{u}|}
    \sum_{x\in\mathcal{X}_{u}}
    \cos(\mathbf{g}_{j}(x),\mathbf{r}_{j})
    -
    \frac{1}{|\mathcal{X}_{s}|}
    \sum_{x\in\mathcal{X}_{s}}
    \cos(\mathbf{g}_{j}(x),\mathbf{r}_{j}).
    \label{eq:gs_gap}
\end{equation}
We retain
\begin{equation}
    \mathcal{J}^{\star}
    =
    \{j\in\mathcal{J}:\Delta_j>\lambda_{\mathrm{gap}}\}
    \label{eq:gs_critical}
\end{equation}
as the safety-critical slice set, where we set $\lambda_{\mathrm{gap}}=1$ following GradSafe-Zero.
Because this phase uses only the server-owned model and reference data, it is performed entirely in plaintext.
We also precompute
\begin{equation}
    \hat{\mathbf{r}}_{j}
    =
    \frac{\mathbf{r}_{j}}{\|\mathbf{r}_{j}\|_2},
    \qquad j\in\mathcal{J}^{\star},
    \label{eq:gs_normalized_reference}
\end{equation}
so that encrypted cosine-similarity evaluation requires homomorphic normalization only for the online gradient $\mathbf{g}_j(x)$.

\noindent
\textbf{Backward Initialization.}
After the encrypted prompt prefill, the resulting state is shared by the target-model decoding and GradSafe detection.
GradSafe constructs the gradient of a causal-LM objective toward a fixed compliance response $r_c$, without modifying the actual decoding sequence.
For a single response token, the backward pass can be initialized directly from the prefill logits as
\begin{equation}
    \frac{\partial\mathcal{L}}{\partial\mathbf{z}}
    =
    \mathbf{p}-\mathbf{e}_{r_c},
\end{equation}
where $\mathbf{p}$ is the softmax output at the last prompt position and $\mathbf{e}_{r_c}$ is the one-hot representation of the compliance token.
The resulting backward pass can proceed independently of subsequent autoregressive decoding.

\noindent
\textbf{Low-Level Checkpointing.}
Efficient backpropagation requires intermediate values from the forward pass.
Keeping all such ciphertexts at their original levels, however, incurs substantial memory consumption because high-level CKKS ciphertexts contain more RNS limbs.
We therefore level-drop the required checkpoints to a designated low storage level immediately after their last forward use.
When a checkpoint is required during backward propagation, we bootstrap it and subsequently level-align it with the current backward gradient ciphertext before evaluation.
This strategy trades additional bootstrapping for substantially lower checkpoint memory consumption.

\noindent
\textbf{Checkpointed Nonlinear Derivatives.}
The nonlinear operations required in the transformer backward pass can be efficiently evaluated by reusing quantities already computed during the forward pass.
For a softmax input $\mathbf{a}$, let $\mathbf{p}=\operatorname{Softmax}_{\mathrm{HE}}(\mathbf{a})$ denote its checkpointed output, and let $\bar{\mathbf{p}}$ denote the incoming gradient during backward propagation.
The gradient with respect to $\mathbf{a}$ is
\begin{equation}
    \bar{\mathbf{a}}
    =
    \mathbf{p}\odot
    \left(
        \bar{\mathbf{p}}
        -
        \langle\bar{\mathbf{p}},\mathbf{p}\rangle\mathbf{1}
    \right).
    \label{eq:gs_softmax_backward}
\end{equation}
Thus, no additional exponential evaluation is required during the softmax backward pass.

For RMSNorm, let $\mathbf{x}\in\mathbb{R}^{d}$ denote its input and define the inverse RMS value
\begin{equation}
    r
    =
    \left(
        \frac{1}{d}\|\mathbf{x}\|_2^2+\varepsilon_{\mathrm{rms}}
    \right)^{-1/2},
\end{equation}
where $\varepsilon_{\mathrm{rms}}$ is the numerical-stability constant of RMSNorm.
Given the RMSNorm weight $\mathbf{w}$ and the incoming gradient $\bar{\mathbf{y}}$ with respect to its output $\mathbf{y}=r\,\mathbf{w}\odot\mathbf{x}$, define
$\mathbf{u}=\bar{\mathbf{y}}\odot\mathbf{w}$.
Using the checkpointed $\mathbf{x}$ and $r$, the input gradient is
\begin{equation}
    \bar{\mathbf{x}}
    =
    r\mathbf{u}
    -
    \frac{r^3}{d}
    \mathbf{x}
    \langle\mathbf{u},\mathbf{x}\rangle.
    \label{eq:gs_rms_backward}
\end{equation}
Hence, the inverse square root already evaluated during the forward pass can be reused without another $\operatorname{InvSqrt}_{\mathrm{HE}}$ evaluation.

For SiLU, let
\begin{equation}
    y=u\,s,
    \qquad
    s=\operatorname{Sigmoid}_{\mathrm{HE}}(u),
\end{equation}
where $s$ and $y$ are checkpointed during the forward pass.
For an incoming gradient $\bar{y}$, the input gradient is
\begin{equation}
    \bar{u}
    =
    \bar{y}
    \left[
        s+y(1-s)
    \right].
    \label{eq:gs_silu_backward}
\end{equation}
Therefore, the sigmoid approximation does not need to be reevaluated during backward propagation.

\noindent
\textbf{Streaming Gradient Similarity.}
To reduce memory usage, we do not store all safety-critical gradients.
Instead, each gradient slice is immediately compared with its pre-normalized unsafe reference using encrypted cosine similarity, accumulated into a running score, and then discarded.
GradSafe-Zero averages the accumulated similarities and applies the online threshold $0.25$ through $\operatorname{sgn}_{\mathrm{HE}}$ to obtain the final acceptance gate.

\subsection{Pre-Processing HE-Guardrails}
\label{sec:pre_guardrails}
Pre-processing guardrails evaluate the prompt independently of target response generation.
We instantiate this approach with Llama Guard 3-8B~\cite{Grattafiori2024Llama}.
Let $x$ be the user prompt and $\mathcal{T}_{\mathrm{T}}$, $\mathcal{T}_{\mathrm{LG}}$ be the target and guardrail prompt templates.
The model inputs are $x_{\mathrm{T}}=\mathcal{T}_{\mathrm{T}}(x)$ and $x_{\mathrm{LG}}=\mathcal{T}_{\mathrm{LG}}(x)$; the latter includes the guardrail's safety-assessment instructions.

For $M\in\{\mathrm{T},\mathrm{LG}\}$, let $n_M$ be the token count and $t_i^M$ the $i$-th token ID after formatting and tokenization.
The corresponding one-hot input and embedding are
\begin{equation}
    \mathbf{o}_i^M=\mathbf{e}_{t_i^M}\in\{0,1\}^{V_M},
    \qquad
    \mathbf{x}_i^M=E_M\mathbf{o}_i^M,
    \label{eq:pre_embedding}
\end{equation}
where $E_M\in\mathbb{R}^{d_M\times V_M}$ is model $M$'s plaintext embedding table.
The two models use their own weights and embedding tables, while the user-dependent one-hot inputs and embeddings remain encrypted.
The formatted streams are padded to a common length with their respective attention masks.

\subsubsection{Multi-Model Serving}
\label{sec:multimodel}
Llama Guard requires an additional HE-LLM evaluation beyond the target model.
Although the two models can run concurrently, executing them as independent HE pipelines doubles the ciphertext memory consumption.
To avoid this overhead, we extend MOAI's interleaved batching~\cite{Zhang2026MOAI} by assigning model-specific plaintext weights to each interleaved lane instead of repeating one model's weights across all lanes, so that the target and guardrail models are evaluated within a single set of ciphertexts.

For lane $b\in\{0,\ldots,B-1\}$, let $X^{(b)}\in\mathbb{R}^{n\times d}$ be its activation and $W^{(b)}\in\mathbb{R}^{d\times d'}$ its weight matrix.
For ciphertexts under a common HE key, the interleaved assignments are
\begin{equation}
\begin{aligned}
    \ct_i[Br+b]&=X^{(b)}[r,i],\\
    \pt_{i,j}[Br+b]&=W^{(b)}[i,j],
\end{aligned}
\label{eq:multimodel_packing}
\end{equation}
for $0\leq r<n$, $0\leq i<d$, and $0\leq j<d'$.
The packed projection is
\begin{equation}
    \ct_j^{\mathrm{out}}=
    \sum_{i=0}^{d-1}\mathsf{Mult}_p(\ct_i,\pt_{i,j}),
    \label{eq:multimodel_linear}
\end{equation}
whose slot $Br+b$ represents $(X^{(b)}W^{(b)})[r,j]$.

\subsubsection{Llama Guard Decision}
\label{sec:llamaguard}
Let $\mathbf{z}_{\mathrm{LG}}\in\mathbb{R}^{V_{\mathrm{LG}}}$ be Llama Guard's logits at the first safety-decision position, and let $t_{\mathrm{safe}}$ and $t_{\mathrm{unsafe}}$ be the corresponding token IDs.
Our implementation blocks when $\mathbf{z}_{\mathrm{LG}}[t_{\mathrm{unsafe}}]>\mathbf{z}_{\mathrm{LG}}[t_{\mathrm{safe}}]$.
We therefore compute the encrypted acceptance margin and gate directly as
\begin{equation}
\begin{aligned}
    m_{\mathrm{LG}}
    &=\mathbf{z}_{\mathrm{LG}}[t_{\mathrm{safe}}]
      -\mathbf{z}_{\mathrm{LG}}[t_{\mathrm{unsafe}}],\\
    g_{\mathrm{LG}}
    &=\frac{1+\operatorname{sgn}_{\mathrm{HE}}(m_{\mathrm{LG}})}{2}.
\end{aligned}
\label{eq:lg_gate}
\end{equation}
This comparison does not require a softmax over the vocabulary or generation of a textual explanation.
It is equivalent to comparing the two token probabilities because Softmax preserves logit ordering.
The gate is passed directly to Eq.~\eqref{eq:gated_response}.

\section{Experiments}
\label{sec:experiments}
Following the Security-Efficiency-Utility (SEU) evaluation framework of Wang et al.~\cite{Wang2026Sok}, we evaluate \hegr along three perspectives: security (attack success rate, pass guardrail rate), efficiency (memory usage and latency overhead), and utility (false rejection of benign prompts).
The target responses and guardrail decisions are computed in the encrypted domain.

\subsection{Experimental Setup and Metrics}
\label{sec:experimental_setup}

\noindent
\textbf{Implementation.}
We implemented all HE models and guardrails using \texttt{desilofhe}~\cite{Desilo2025library}, a Python-based HE library that supports the RNS-CKKS scheme with GPU acceleration.
We set the number of slots as $N/2=2^{15}$ and the available multiplicative level after bootstrapping as $17$.
All experiments were conducted on two NVIDIA H200 GPUs running AlmaLinux~9.8: the target model Llama-3-8B-Instruct occupies one GPU, and a separately scheduled guardrail occupies the other.
For the target model, we prioritize numerical accuracy over depth: SiLU is approximated by a degree-127 polynomial, RMSNorm evaluates the inverse square root via Goldschmidt iterations, and Softmax follows Cho et al.~\cite{Cho2024Fast}.

The padded input length is 2,048 tokens, including any guardrail instructions, and the generation budget is 200 tokens.
Under multi-model serving, this yields $2^{15}/2{,}048=16$ interleaved lanes.
Joint target--Llama Guard execution assigns two lanes per request, so the same layout admits eight guarded requests instead of sixteen target-only requests.

\noindent
\textbf{Attack and Benign Inputs.}
Following Wang et al.~\cite{Wang2026Sok}, we evaluate six jailbreak attack families comprising 2,215 prompts or sessions: IJP~\cite{Shen2024Do} (manual), GCG~\cite{Zou2023Universal} and AutoDAN~\cite{Liu2024Autodan} (optimization-based), DrAttack~\cite{Li2024DrAttack} and MultiJail~\cite{Deng2024Multilingual} (implicit), and ActorAttack~\cite{Ren2025LLMs} (multi-turn).
For benign-input evaluation, we use AlpacaEval~\cite{Li2023AlpacaEval} and OR-Bench~\cite{Cui2025OR} to measure false-positive rejection.

\noindent
\textbf{Metrics and Judge.}
We judge target responses with \texttt{cais/HarmBench-Llama-2-13b-cls}~\cite{Mazeika2024HarmBench}.
For an attack sample $i$, let $a_i\in\{0,1\}$ indicate guardrail acceptance and $h_i\in\{0,1\}$ indicate a harmful-response judgment.
For an attack set $\mathcal{A}$, attack success rate (ASR) and pass-guardrail rate (PGR) are
\begin{equation}
    \mathrm{ASR}=\frac{\sum_{i\in\mathcal{A}}a_i h_i}{|\mathcal{A}|},
    \qquad
    \mathrm{PGR}=\frac{\sum_{i\in\mathcal{A}}a_i}{|\mathcal{A}|}.
    \label{eq:attack_metrics}
\end{equation}
For a benign set $\mathcal{B}$, the false-positive rate is $\mathrm{FPR}=|\mathcal{B}|^{-1}\sum_{i\in\mathcal{B}}(1-a_i)$.
PGR separates prompt rejection from the target's own refusal behavior: low ASR alone need not indicate an effective guardrail.

\begin{table*}[t]
\centering
\caption{Attack success rate (ASR) and pass-guardrail rate (PGR) in percent; lower is better.
All target responses and guardrail decisions are computed in the encrypted domain, and harmfulness is judged by \texttt{HarmBench-Llama-2-13b-cls}~\cite{Mazeika2024HarmBench}.
}
\label{tab:attack_results}
\small
\setlength{\tabcolsep}{6pt}
\begin{tabular}{lrrrrrrrr}
\toprule
& & No defense & \multicolumn{2}{c}{Llama Guard 3-8B} & \multicolumn{2}{c}{JBShield-D} & \multicolumn{2}{c}{GradSafe-Zero}\\
\cmidrule(lr){4-5}\cmidrule(lr){6-7}\cmidrule(lr){8-9}
Attack & Samples & ASR & ASR & PGR & ASR & PGR & ASR & PGR\\
\midrule
\multicolumn{9}{l}{\emph{Single-turn attacks}}\\
IJP          & 1,000 & 14.3 & 9.6 & 33.6 & 14.3 & 95.4 & 14.2 & 59.9\\
GCG          &   100 & 13.0 & 2.0 &  3.0 & 13.0 & 72.0 & 13.0 & 77.0\\
AutoDAN      &   100 &  2.0 & 0.0 & 21.0 &  2.0 & 46.0 &  1.0 &  4.0\\
DrAttack     &   100 & 11.0 & 7.0 & 39.0 & 11.0 & 52.0 & 11.0 & 42.0\\
MultiJail    &   315 &  9.5 & 7.9 & 37.1 &  9.5 & 97.1 &  9.5 & 91.1\\
\midrule
\multicolumn{9}{l}{\emph{Multi-turn attacks}}\\
ActorAttack  &   600 & 16.3 & 16.2 & 98.5 & 16.3 & 99.2 & 16.2 & 99.8\\
\midrule
Average  & -- & 11.0 & 7.1 & 38.7 & 11.0 & 77.0 & 10.8 & 62.3\\
\bottomrule
\end{tabular}
\end{table*}

\subsection{Jailbreak Defense}
\label{sec:asr_results}
Across all evaluated prompts, the encrypted guardrails reproduce the plaintext decisions in 99.7\% of cases, indicating that the polynomial approximations and HE numerical error rarely alter the safety decision itself.
The following comparisons therefore reflect the underlying detection rules rather than HE-induced degradation.

Table~\ref{tab:attack_results} reports ASR for the six evaluated attack families.
The macro-average ASR is 11.0\% without a guardrail, 7.1\% for Llama Guard, 11.0\% for JBShield, and 10.8\% for GradSafe.
Llama Guard lowers ASR on every family, most notably from 13.0\% to 2.0\% on GCG and from 14.3\% to 9.6\% on IJP, while ActorAttack changes only from 16.3\% to 16.2\%.
JBShield matches the no-defense ASR on all six families, and GradSafe differs by at most 1.0 percentage point.

The average PGR is 38.7\% for Llama Guard, 77.0\% for JBShield, and 62.3\% for GradSafe.
All three guardrails pass 98.5--99.8\% of ActorAttack sessions.
The per-family values vary widely: on GCG, PGR is 3.0\% for Llama Guard but 72.0\% for JBShield and 77.0\% for GradSafe, whereas on AutoDAN it is 21.0\%, 46.0\%, and 4.0\%, respectively.

\subsection{Memory Usage and Latency Overhead}
\label{sec:efficiency}
\noindent
\textbf{GPU Memory Usage.}
Table~\ref{tab:memory} reports approximate memory increases relative to running the target model alone.
A separately evaluated Llama Guard adds approximately 140\,GiB and occupies the second H200 GPU.
JBShield adds approximately 5\,GiB because its online work uses a hidden-state vector and calibrated references.
GradSafe adds approximately 110\,GiB: streaming similarities avoid retaining all gradients, but backward computation still requires a substantial checkpoint working set.

Joint target--Llama Guard serving incurs no additional ciphertext memory, since both models share the same interleaved ciphertexts.
This comes at the cost of halved batch capacity: the sixteen lanes accommodate eight guarded requests instead of sixteen target-only requests.

\begin{table}[t]
\centering
\caption{Approximate incremental GPU memory relative to target-only inference.
}
\label{tab:memory}
\small
\setlength{\tabcolsep}{4pt}
\begin{tabular}{lr}
\toprule
Configuration & Additional memory (GiB)\\
\midrule
Llama Guard: separate execution & $\approx140$\\
Llama Guard: multi-model serving & $\approx0^{*}$\\
JBShield-D & $\approx5$\\
GradSafe-Zero & $\approx110$\\
\bottomrule
\end{tabular}
\par\smallskip
\begin{minipage}{\columnwidth}
\footnotesize
$^{*}$At fixed sixteen-lane capacity: eight target--guardrail pairs instead of sixteen target-only requests.
\end{minipage}
\end{table}

\noindent
\textbf{Concurrent Latency.}
All three guardrails run concurrently with target-model decoding.
Llama Guard is evaluated alongside the target prefill under multi-model serving, whereas JBShield and GradSafe start as soon as the prefill signals they require become available and proceed independently of subsequent decoding.
Following the SEU framework~\cite{Wang2026Sok}, let $\tau_{\mathrm{T}}$ denote the end-to-end response time of target-only inference and $\tau_{\mathrm{T+G}}$ that of the target model with a guardrail, both measured from request submission to the release of the final response.
The latency overhead of the guardrail is
\begin{equation}
    \Delta\tau
    =\tau_{\mathrm{T+G}}-\tau_{\mathrm{T}},
    \label{eq:guardrail_wait}
\end{equation}
which is zero for every evaluated guardrail in our two-GPU configuration: because the 200-token autoregressive decoding dominates end-to-end latency, each guardrail decision is ready before the final response is released.
The guardrail therefore adds no latency on the target model's critical path, although it still consumes computation and memory on the second GPU.

\subsection{Benign-Prompt Utility}
\label{sec:benign_utility}
Table~\ref{tab:fpr} reports the FPR of each guardrail on benign prompts.
Llama Guard rejects 33 of 1,805 benign prompts (1.83\%), compared with one (0.06\%) for JBShield and three (0.17\%) for GradSafe.
On OR-Bench, which is designed to probe over-refusal, Llama Guard rejects 22 of 1,000 prompts.
Its stronger attack rejection therefore comes at the cost of higher benign rejection, whereas the low FPRs of JBShield and GradSafe reflect their more permissive detection rules rather than uniformly better guardrails.

\begin{table}[t]
\centering
\caption{False-positive rate (\%) on benign prompts, with the number of rejected prompts in parentheses.}
\label{tab:fpr}
\small
\setlength{\tabcolsep}{4pt}
\begin{tabular}{lrrr}
\toprule
Guardrail & \shortstack{AlpacaEval\\($n=805$)} & \shortstack{OR-Bench\\($n=1{,}000$)} & \shortstack{Combined\\($n=1{,}805$)}\\
\midrule
Llama Guard & 1.37 (11) & 2.20 (22) & 1.83 (33)\\
JBShield-D & 0.12 (1) & 0.00 (0) & 0.06 (1)\\
GradSafe-Zero & 0.12 (1) & 0.20 (2) & 0.17 (3)\\
\bottomrule
\end{tabular}
\end{table}

\section{Future Work}
Multi-turn jailbreak attacks remain an important limitation of current guardrail mechanisms.
As reported by Wang et al.~\cite{Wang2026Sok}, even session-level guardrails exhibit substantial vulnerability to multi-turn attacks, particularly against adaptive attacks such as X-Teaming~\cite{Rahman2025XTeaming}.
This limitation arises because the malicious intent may emerge gradually across multiple interactions rather than being identifiable from an individual prompt or response.

In plaintext LLM deployments, conversation histories can at least be directly inspected by service operators or subjected to human moderation as an operational fallback.
In HE-LLM inference, however, the server cannot directly inspect either the prompts or generated responses, making such plaintext-level intervention unavailable.
Given both the high attack success rates of multi-turn jailbreaks and the lack of plaintext visibility under HE, developing HE-compatible guardrails that securely reason over encrypted conversation histories remains an important direction for future work.

\section{Conclusion}
\label{sec:conclusion}
In this work, we identified a security vulnerability in HE-LLM inference: the confidentiality that protects benign clients also allows malicious clients to submit jailbreak prompts that the server can neither inspect nor block.
To address this vulnerability, we proposed \hegr, a framework that evaluates jailbreak guardrails entirely over encrypted data and enforces their decisions through homomorphic response gating with gated noise flooding, so that the suppressed response cannot be recovered by the client.
We instantiated \hegr with Llama Guard, JBShield, and GradSafe, covering both pre-processing and intra-processing guardrails, and developed the HE-compatible building blocks they require, including encrypted gradient evaluation for GradSafe and multi-model serving for concurrent evaluation of the target and guardrail models.
Our evaluation shows that the encrypted guardrails achieve defense performance close to that of their plaintext counterparts, with distinct security--efficiency--utility trade-offs across guardrail designs, and that guardrail evaluation can be overlapped with target generation without adding latency to the response.

More broadly, our threat model departs from the conventional assumption in HE-based PPML that the client is benign in its choice of input.
Considering a client that follows the protocol yet adversarially exploits the encrypted model is a necessary step toward trustworthy HE-based inference, in which the model itself must be protected in addition to the client's data.
Overall, these results bring HE-LLM inference closer to practical deployment and open a direction for further research on safety mechanisms for encrypted LLM inference.





\bibliographystyle{IEEEtran}
\bibliography{abrv,mybib,mybib_jailbreak}
%




\end{document}